\documentclass[twocolumn,prl,showpacs]{revtex4}

\usepackage{graphicx}
\usepackage{amssymb}
\usepackage{amsfonts}

\DeclareGraphicsExtensions{.eps}

\begin{document}
\title{Polarization squeezing with cold atoms}

\date{\today}

\author{V. Josse}

\author{A. Dantan}

\author{L. Vernac}

\author{A. Bramati}

\author{M. Pinard}

\author{E. Giacobino}

\affiliation{Laboratoire Kastler Brossel, Universit\'{e} Pierre et
Marie Curie , Ecole
Normale Sup\'{e}rieure et CNRS,\\
Case 74, 4 place Jussieu, 75252 Paris Cedex 05, France}

\begin{abstract}
We study the interaction of a nearly resonant linearly polarized
laser beam with a cloud of cold cesium atoms in a high finesse
optical cavity. We show theoretically and experimentally that the
cross-Kerr effect due to the saturation of the optical transition
produces quadrature squeezing on both the mean field and the
orthogonally polarized vacuum mode. An interpretation of this
vacuum squeezing as \textit{polarization squeezing} is given and a
method for measuring quantum Stokes parameters for weak beams via
a local oscillator is developed.

\end{abstract}


\newcommand{\rf}[1]{(\ref{#1})}
\newcommand{\beq}{\begin{equation}}
\newcommand{\eeq}{\end{equation}}
\newcommand{\beqr}{\begin{eqnarray}}
\newcommand{\eeqr}{\end{eqnarray}}
\newcommand{\lb}[1]{\label{#1}}
\newcommand{\ct}[1]{\cite{#1}}
\newcommand{\bi}[1]{\bibitem{#1}}
\newcommand{\bk}{_{\bf k}}

\maketitle
\newpage

A great deal of attention has been recently given to the quantum
features of the polarization states of the light, essentially
because of their connections with quantum information technology.
Several theoretical schemes to produce polarization squeezing
using Kerr-like media have been proposed \ct{Chirkin} and realized
using optical fibers \ct{Boivin}. Other experimental realizations
achieve polarization squeezing by mixing squeezed vacuum
(generated by an OPO) with a strong coherent beam on a polarizing
beam splitter \ct{Grangier} or mixing two independent quadrature
squeezed beams (generated by an OPA) on a polarizing beam splitter
\ct{Bowen}. Very recently it has been proposed to propagate a
linearly polarized light beam through an atomic medium exhibiting
self rotation to generate squeezed vacuum in the orthogonal
polarization \ct{Matsko}, which is equivalent to
achieving polarization squeezing.\\
In previous works \ct{Lambrecht} the interaction between a cloud
of cold cesium atoms placed in a high finesse optical cavity and a
\textit{circularly} polarized laser beam nearly resonant with an
atomic transition has been studied. Because of optical pumping,
the atomic medium is conveniently modelled by an ensemble of
two-level atoms. The saturation of the optical transition gives
rise to an intensity-dependent refraction index. It is well known
that the interaction of the light with a Kerr-like medium produces
bistable behavior of the light transmitted by the cavity and that,
at the turning point of the bistability curve, the quantum
fluctuations of the light can be strongly modified and generate
quadrature squeezing \ct{hilico}. A noise reduction of 40\% has
thus been
observed in our group \ct{Lambrecht}.\\
In this paper we focus on the theoretical and experimental
investigation of polarization squeezing via the interaction of a
\textit{linearly} polarized laser beam with cold cesium atoms. In
this configuration, the two-level atom model is no longer
applicable and the situation much more complicated. We describe
the interaction between light and the atomic medium by means of an
X-like four-level quantum model based on the linear input-output
method. Our theoretical analysis shows clearly that competitive
optical pumping may result in polarization switching, and
polarization squeezing is predicted by the model \ct{Josse}. In
agreement with the model we observe quadrature squeezing in the
probe laser mode \textit{and} in the orthogonal vacuum mode.
Experimentally, we obtain a polarization squeezing of $13\%$ and
we show for the first time that the quantum Stokes parameters for
very weak beams can be measured together with their phases using a
local oscillator (LO). In our case squeezing is due to cross-Kerr
effect induced by the mean field rather than to the polarization
self-rotation
responsible for polarization switching.\\

\begin{figure}[h]
\includegraphics[scale=1.05]{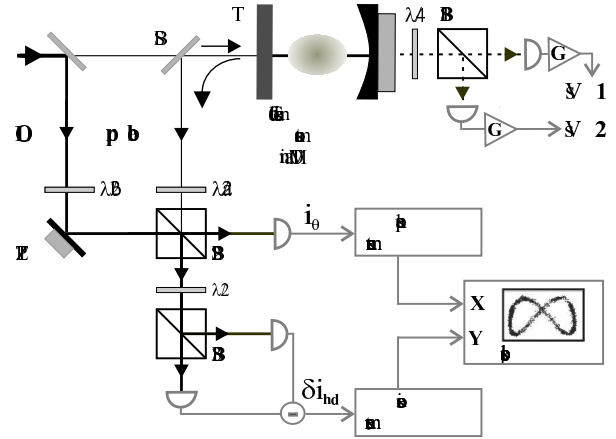}
\caption{Experimental set-up: PBS: polarizing beam splitter; BS:
10/90 beam splitter; $\lambda/2$: half-wave plate; PZT:
piezo-electric ceramic.} \label{fig2}
\end{figure}

The configuration used in the experiment is described in detail in
\ct{Lambrecht}. In Fig. 1 we present the main features of the
set-up. The cesium atoms are cooled in a standard magneto-optical
trap which operates with three orthogonal circularly polarized
trapping beam generated by a Ti:Sapphire laser and an
inhomogeneous magnetic field. The trapping Ti:Sapphire laser is
detuned by 3 times the linewidth of the upper state on the low
frequency side of the 6S$_{1/2}$, F=4 to 6P$_{3/2}$, F=5
transition. To prevent atoms from being optically pumped to the
6S$_{1/2}$, F=3 state, we superimpose a diode laser tuned to the
6S$_{1/2}$, F=3 to 6P$_{3/2}$, F=4 transition to the trapping
beams. We use a 25 cm long linear cavity built around the cell.
The cavity is close to the hemifocal configuration with a waist of
260 $\mu$m. The coupling mirror has a transmission coefficient $T$
of 10\%, the rear mirror is highly reflecting. Hence, the cavity
is close to the bad cavity limit for which the cavity linewidth
($\kappa=5$ MHz) is larger than the atomic linewidth ($\gamma=2.6$
MHz). We probe the atoms with a linearly polarized laser beam
detuned by about 50 MHz in the red of the 6S$_{1/2}$, F=4 to
6P$_{3/2}$, F=5 transition. The optical power of the probe beam
ranges from 5 to 15 $\mu$W. Under these conditions, the number of
atoms interacting with the light beam is about $10^6$-$10^7$. The
polarization of the light transmitted by the cavity is analyzed
with a quarter-wave plate and a polarizing beamsplitter (PBS1), so
that Vs1 and Vs2 give the amount of respectively left-handed
($I_+$) and right-handed ($I_-$) circular light.

\begin{figure}[h]
\includegraphics[scale=0.4]{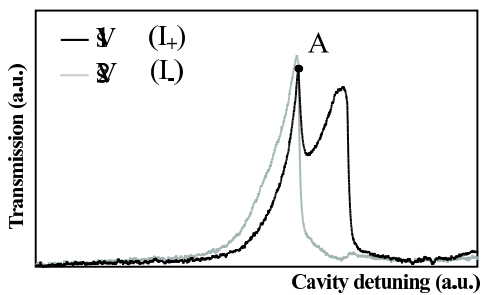}
\caption{Analysis of the circular polarization content of the
light transmitted by the cavity and detected by the photodiodes
shown in Fig. 1. Polarization switching occurs at point $A$. The
power of the incident light is 7 $\mu$W.} \label{fig3}
\end{figure}

A typical recording of the transmitted intensities $I_{\pm}$ as a
function of the cavity length is shown in Fig. 2: starting from
the left up to point $A$, the polarization remains linear (nearly
equal amount of circular polarized light on both photodiodes),
then it becomes circular. This polarization switching was known to
occur in Fabry-Perot cavities containing atomic vapors with
degenerate sublevels in the ground state \ct{Cecchi,Giacobino}. In
the experiment the probe beam is nearly resonant with the
6S$_{1/2}$, F=4 to 6P$_{3/2}$, F=5 transition: in principle, all
the 20 Zeeman sub-levels are involved in the interaction. In order
to get a qualitative physical insight into this problem, avoiding
too heavy calculations, we model the atomic medium as an $X$-like
four-level system [see Fig. 3].

\begin{figure}[h]
\includegraphics[scale=0.3]{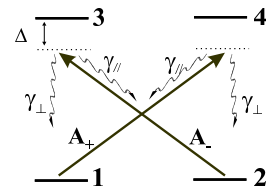}
\caption{Schematic energy level diagram for the X-like four-level
system: $\gamma_{\perp}+\gamma_{\parallel}=\gamma$ is the optical
dipole decay rate; $\Delta$ is the (large) detuning from
resonance.}\label{fig1}
\end{figure}

The competitive optical pumping between the circular component
$\sigma_{\pm}$ of light makes the linear polarization unstable
inside the cavity above some intensity threshold. The optical
pumping is responsible for polarization switching and the general
shape of the cavity resonance curve [see Fig. 2] is well
understood within this theoretical frame
\ct{Josse,Giacobino,Cecchi}. Alternatively, the polarization
switching threshold may be interpreted as a laser oscillation
threshold for the mode orthogonal to the main polarization mode
\ct{Josse}. Here, we are interested in the quantum fluctuations of
the light, which can be strongly modified via the interaction with
the atoms. When the polarization of the light is circular, the
situation is analogous to the previous experimental scheme when
the incoming field was circularly polarized \ct{Lambrecht}. We
will therefore focus in the following on the case for which the
polarization remains linear along the $x$-axis. The saturation of
the $\sigma_{\pm}$ components of light causes the medium to behave
as a Kerr-like medium for the mean field $\hat{A}_x$. In addition,
the vacuum orthogonal $\hat{A}_y$ mode experiences a non linear
dephasing via cross-Kerr effect \ct{Boivin,Josse}. This system is
then expected to generate quadrature squeezing for both modes; in
the large detuning limit [$\Delta\gg\gamma$], the equation for the
vacuum mode fluctuations reads \ct{Josse}

\beqr\frac{d\delta\hat{A}_y}{dt}&=&-[\kappa+i(\Delta_c -
\Delta_0)] \delta \hat{A}_y\\
\nonumber &&-i \Delta_{0}\frac{s_{x}}{2}[2 \delta
\hat{A}_{y}-\frac{\langle \hat{A}_{x} \rangle ^{2}}{| \langle
\hat{A}_{x} \rangle|^2} \delta \hat{A}_{y}^{\dag}]+
\frac{2\kappa}{\sqrt{T}} \delta \hat{A}^{in}_{y}\eeqr

with $\Delta_c$ the cavity detuning, $\Delta_0=2Ng^2\kappa/\Delta
T$ the linear atomic dephasing,
$s_x=2g^2|\langle\hat{A}_x\rangle|^2/\Delta^2$ the saturation
parameter and $\delta \hat{A}_y^{in}$ the incident field
fluctuations. The cross-Kerr induced term has two contributions: a
dephasing ($\propto |\langle \hat{A}_x \rangle ^2\
\delta\hat{A}_y$) and the term in $\langle \hat{A}_x \rangle ^2
\delta\hat{A}^{\dag}_y$, responsible for the squeezing of this
mode. A similar equation can be derived for the mean field
$\hat{A}_x$, for which the squeezing is then generated via the
usual Kerr term ($\propto \langle \hat{A}_x \rangle ^2
\delta\hat{A}^{\dag}_x$). These equations are valid when the large
excess noise due to optical pumping can be neglected, that is for
times smaller than the optical pumping time, in contrast with
\ct{Matsko}. Squeezing is then predicted for noise frequencies
higher than the inverse optical pumping
time.\\
Using the experimental set-up described in Fig. 1, we have
measured the quadrature noise of both modes. The signal recorded
at the output of the cavity (see Fig. 2) is used to lock the
cavity length on the regime where the polarization remains linear.
After interacting with the atoms, both fields are mainly reflected
by BS and then mixed on PBS2 with a 10 mW local oscillator beam
(LO). Using the half-wave plate $\lambda/2a$, we are able to send
either the mean field mode or the orthogonal vacuum mode to PBS3
and perform the usual homodyne detection. By varying the relative
phase of the LO with respect to the probe beam we can detect the
noise features of all the quadratures of the field. In agreement
with the theoretical predictions, we observe quadrature squeezing
on both the main mode and the orthogonal mode. The results are 5\%
(7\% after correction for optical losses, mainly due to BS) of
noise reduction for the main polarization mode at 6 MHz and 13\%
(18\% after correction) for the squeezed vacuum state at 3 MHz, as
shown in Fig. 4. This system is then able to produce
simultaneously two squeezed modes, for which the relatives phases
are intrinsically fixed. Let us note that these two squeezed modes
can be used to generate a pair of entangled beams, which will be
presented in a forthcoming publication \ct{Josse2}.

\begin{figure}[h]
\includegraphics[scale=0.4]{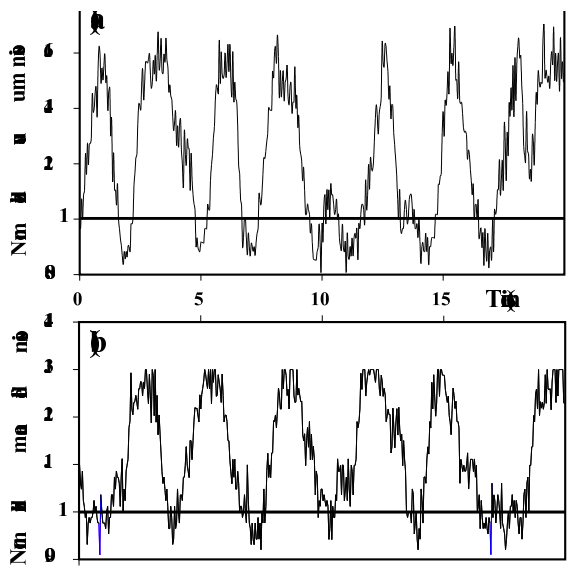}
\caption{Normalized quadrature noise for the vacuum field mode at 3 MHz (a)
and for the mean
field mode at 6 MHz (b). The best squeezing is $13\%$ for the vacuum mode
and $5\%$ for the mean
field mode.}\label{fig4}
\end{figure}

In the following we focus on the study of the noise of the mode
with orthogonal polarization with respect to the main mode,
commonly referred to as \textit{polarization noise}. The
characterization of the quantum features of the polarization state
relies on the measurement of the quantum Stokes parameters
\ct{Korolkova}. They are defined from their classical counterparts

\beqr \hat{S}_{0}&=&\hat{A}^{\dag}_{x}\hat{A}_{x}+\hat{A}^{\dag}_{y}\hat{A}_{y}
\;,\hspace{0.5cm}
\hat{S}_{1}=\hat{A}^{\dag}_{x}\hat{A}_{x}-\hat{A}^{\dag}_{y}\hat{A}_{y}\nonumber\\
\hat{S}_{2}&=&\hat{A}^{\dag}_{x}\hat{A}_{y}+\hat{A}^{\dag}_{y}\hat{A}_{x}\;,
\hspace{0.5cm}\hat{S}_{3}=i(\hat{A}^{\dag}_{y}\hat{A}_{x}-\hat{A}^{\dag}_{x}\hat{A}_{y})
\nonumber
\eeqr

These operators obey the commutation relations $
[\hat{S}_{0},\hat{S}_{i}]=0\hspace{0.2cm}$ and
$[\hat{S}_{i},\hat{S}_{j}]=\epsilon_{ijk}2i\hat{S}_{k}$
($i=1,2,3$). Their spectral noise densities satisfy uncertainty
relations $V_{\hat{S}_{i}}(\omega)V_{\hat{S}_{j}}(\omega)\geq |
\epsilon_{ijk}\langle\hat{S}_{k}\rangle|^2$. In our case the light
is linearly polarized along the $x$-axis, then $\langle \hat{S}_0
\rangle =\langle \hat{S}_1 \rangle =\alpha_x^2$ and $\langle
\hat{S}_2 \rangle =\langle \hat{S}_3 \rangle =0$, where $ \langle
\hat{A}_x \rangle = \alpha_x $ is chosen real. Then the only non
trivial Heisenberg inequality  is
$V_{\hat{S}_{2}}(\omega)V_{\hat{S}_{3}}(\omega)\geq\alpha_x^4$.
\textit{Polarization squeezing} is then achieved if
$V_{\hat{S}_{2}}$ or $V_{\hat{S}_{3}}$ is below the coherent state
value $\alpha_x^2$. The fluctuations of $\hat{S}_2$ and
$\hat{S}_3$ are related to the fluctuations of the quadratures of
the vacuum orthogonal mode

\beqr \delta
\hat{S}_{2}=\alpha_{x}(\delta\hat{A}^{\dag}_{y}+\delta\hat{A}_{y})\;,\;\;
\delta\hat{S}_{3}=i\alpha_{x}(\delta\hat{A}^{\dag}_{y}-\delta\hat{A}_{y})
\label{stokes}\eeqr

The physical meaning of the Stokes parameters fluctuations is the
following: the $\delta\hat{S}_2$ fluctuations lead to a geometric
jitter of the polarization axis, whereas the $\delta\hat{S}_3$
fluctuations are linked to ellipticity fluctuations. It can be
seen from Eq. (\ref{stokes}) that these fluctuations are related
to the amplitude and phase fluctuations of $\hat{A}_y$. Therefore
polarization squeezing is equivalent to vacuum squeezing on the
orthogonal mode.

The measurement of the Stokes parameters can be carried out
directly by means of two balanced photodiodes and suitable
combinations of half-wave and quarter-wave plates \ct{Korolkova}.
In our set-up, however, the power of the probe beam interacting
with the atoms ($\sim 10\mu$W) is not sufficient, so that we need
a LO for the detection. The fluctuations of the vacuum mode
$\hat{A}_y$ are measured using the homodyne detection described
above. Following Eq. (\ref{stokes}) the photocurrent can be
expressed in terms of the fluctuations of $\hat{S}_2$ and
$\hat{S}_3$:

\beq \delta i_{hd}\propto\cos\theta_{hd}\delta\hat{S}_{2}
+\sin\theta_{hd}\delta\hat{S}_{3}\equiv
\delta\hat{S}_{\theta_{hd}}\label{idh} \eeq

where $\theta_{hd}$ is the relative phase between the LO and the
mean field. As $\theta_{hd}$ is varied in time, we correspondingly
detect the fluctuations of the Stokes parameter $
\hat{S}_{\theta_{hd}}$. For instance, $\theta_{hd}=0$
(respectively $\pi/2$) corresponds to the detection of the
fluctuations of $\hat{S}_{2}$ (respectively of $\hat{S}_{3}$).
Hence, in the experiment we can get the Stokes parameters simply
by simultaneously measuring the relative phase $\theta_{hd}$ and
the quadrature noise of the vacuum mode. This measurement is
readily performed by setting the half-wave plate before PBS2 in
such a way that the $\hat{A}_{y}$ mode is sent to the homodyne
detection; the mean field $\hat{A}_{x}$ goes through the other
port of the beam splitter and is detected together with a portion
of the LO by a photodiode (see Fig. 1). The phase is determined
via the interference signal between LO and $\hat{A}_x$
($i_{\theta}\propto \cos \theta_{hd})$. The two signals
$i_{\theta} $ and $\delta i_{hd}$ are sent to the $XY$ channel of
the oscilloscope, giving the characteristic curves reported below.

\begin{figure}[h]
\includegraphics[scale=0.5]{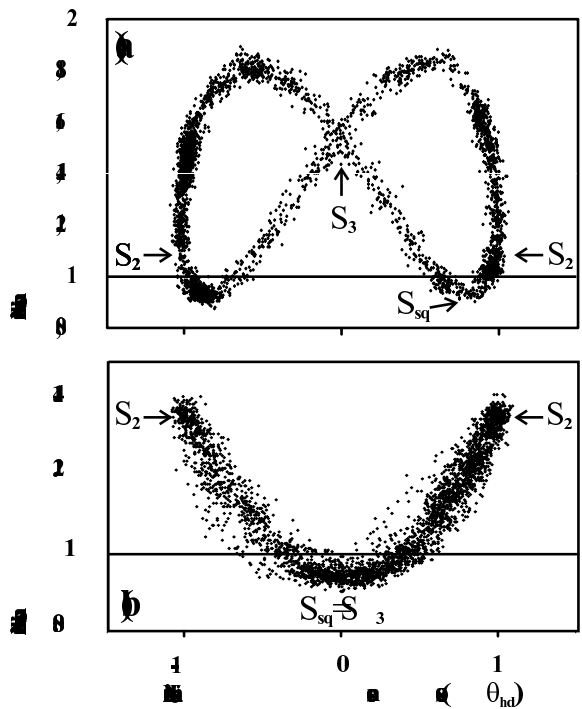}
\caption{Normalized quadrature noise at 3 MHz for the vacuum mode
$\hat{A}_{y}$ \textit{vs} the normalized interference signal:
$\cos \theta_{hd}$. The general case is shown in curve (a):
polarization squeezing is achieved when
$\theta_{hd}=\theta_{sq}=\pm 30^{\circ}$: a linear combination of
$\hat{S}_{2}$ and $\hat{S}_{3}$ is squeezed. In curve (b),
$\hat{S}_{3}$ is squeezed ($\theta_{sq}=\pm \pi/2$).}\label{fig5}
\end{figure}

In Fig. 5 the normalized quadrature noise of $\hat{A}_{y}$,
obtained at a noise frequency of 3 MHz, is plotted as a function
of the relative phase between the mean field and the LO. In
agreement with Eq. (\ref{idh}), it can be seen that the noise of
$\hat{S}_2$ is given by the extreme points $\theta_{hd}=0,\pm\pi$
on the diagram and that of $\hat{S}_3$ by the center point
$\theta_{hd}=\pm\pi/2$. In general, for an arbitrary squeezed
quadrature, a linear combination of $\hat{S}_{2}$ and
$\hat{S}_{3}$ is squeezed (Fig. 5a). We find that the polarization
squeezing strongly depends on the operating
point and on the noise frequency. For instance, in Fig. 5b, we see that $\hat{S}_3$ is squeezed.\\

To conclude, we have demonstrated that the nearly resonant
interaction of a linearly polarized laser beam with a cloud of
cold atoms in an optical cavity can produce quadrature squeezing
on the mean field mode \textit{and} on the orthogonally polarized
vacuum mode. We have shown that these results can be interpreted
as polarization squeezing and developed a method to measure the
quantum Stokes parameters for weak beams, using a local oscillator
and a standard homodyne detection.

\begin{acknowledgments}
This work was supported by the QIPC European Project No.
IST-1999-13071 (QUICOV).
\end{acknowledgments}

\end{document}